\documentclass[aps,prc,reprint,amsmath,superscriptaddress,nofootinbib]{revtex4-1}

\usepackage{color}
\definecolor{theblue}{RGB}{0,50,230}

\usepackage{hyperref}
\hypersetup{
  colorlinks=true,
  linkcolor=theblue,
  citecolor=theblue,
  urlcolor=theblue
}

\usepackage{graphicx}
\graphicspath{{fig/}}

\newcommand{\R}{\mathcal{R}}

\DeclareMathOperator\erf{erf}

\begin{document}

\title{Predicting outcomes for games of skill by redefining what it means to win}

\author{J.\ Scott Moreland}
\affiliation{Department of Physics, Duke University, Durham, NC 27708-0305}
\author{Matthew C.\ Superdock}
\affiliation{Department of Mathematical Sciences, Carnegie Mellon University, Pittsburgh, PA 15213}

\date{\today}

\begin{abstract}
The Elo rating system is a highly successful ranking algorithm for games of skill where, by construction, one team wins and the other loses.
A primary limitation of the original Elo algorithm is its inability to predict information beyond a match's win-loss probability.
Specifically, the victor is awarded the same point bounty if he beats a team by 1 point or 10 points; only the rating difference between the team and its opponent affects the match bounty.
  In this work, we explain that Elo ratings and predictions can be naturally extended to include margin-of-victory information by simply redefining \mbox{``what it means to win.''}
We create ratings for each value of the margin-of-victory and use these ratings to predict the \emph{full} distribution of point spread outcomes for matches which have not yet been played.
\end{abstract}

\maketitle

\section{Elo rating system}

The Elo system ascribes a power ranking index $\R$ to each competitor (player or team) which can be used to predict the outcome of a future matchup \cite{elo1978rating}.
Teams with a higher rating are viewed as more likely to win their matchup, and teams with a lower rating are more likely to lose.
The process is fundamentally Bayesian and rating values before and after the match reflect prior and posterior knowledge of each team's performance.

The convention is to initialize all team ratings at ${\R_0 = 1500}$, although this choice is arbitrary.
As we show momentarily, only the rating difference between two teams matters, and hence the Elo ratings can be initialized with any starting value.
The ratings of the competitors are then updated iteratively after each matchup.
Teams gain or lose points based on the outcome of the match and the rating difference between the two competitors.
One is awarded more points for beating a stronger opponent and less points for beating a weaker opponent.
The points awarded to the victor are deducted from the loser such that the total number of points in the system is always conserved.

More specifically, the victor bounty scales with the degree to which a team exceeds (or falls short of) their expected match performance.
Let $P_\text{exp}$ denote the expected probability that a team beats their opponent at a neutral venue.
The bounties that the team and their opponent receive from the outcome of the match are
\begin{align}
  \label{elo}
  \Delta \R_\text{team} &= \kappa \,(P_\text{obs} - P_\text{exp}),\\
  \Delta \R_\text{opp} &= -\Delta \R_\text{team},
\end{align}
where ${P_\text{obs}=1}$ if the team beats their opponent and ${P_\text{obs}=0}$ if the team loses, while $P_\text{obs}=0.5$ is often used if both teams tie.
The free parameter $\kappa$ modulates the relative weight of prior and posterior information. 
If $\kappa$ is small, the prior holds significant weight, and it takes many successive wins or losses to move the rating up or down.
Conversely, when $\kappa$ is large, each win or loss can significantly affect the team's rating.
Typically, the hyper-parameter $\kappa$ is tuned to optimize the veracity of model predictions by back-testing on historical games.

The expected probability $P_\text{exp}$ that a team beats an opponent is modeled by a normal distribution,\footnote{Both normal and logistic distributions are commonly used in Elo ratings. For the purposes of this work, the difference between the two distributions is negligible.}
\begin{align}
  \label{win_prob}
  P_\text{exp}(\Delta \R) &= \int\limits_{-\infty}^{\Delta \R} \frac{1}{\sqrt{2\pi \sigma^2}} \exp \left( -\frac{{\Delta\R}^2}{2 \sigma^2}\right),\\[1ex]
  &=\frac{1}{2}\left[1 + \erf \left( \frac{\Delta\R}{\sqrt{2} \sigma} \right) \right],
\end{align}
where $\Delta \R = \R_\text{team} - \R_\text{opp}$, and the free parameter $\sigma$ sets the scale of rating differences.
Much like the starting Elo, the parameter $\sigma$ can take any value and only the ratio $\kappa/\sigma$ is meaningful for observable quantities.
We fix the parameter $\sigma=300$ such that an underdog beating a team with a $\Delta\R=300$ point rating advantage occurs $\sim\!16\%$ of the time.

A primary feature of the Elo system is that the expected win probability $P_\text{exp}$, determined from iteratively updated Elo ratings $\R_\text{team}$ and $\R_\text{opp}$, converges on the \emph{true} win probability of the match when the competitors sample scores from independent random variables.
We demonstrate this convergence property of the Elo algorithm by constructing a simple toy model where the underlying point distribution of every team is known.

Consider, for instance, a league of nine different teams which each sample $k$ points from a Poisson distribution $P(k; \lambda$) using one of nine different means,
\begin{equation}
  \label{teams}
  \lambda_\text{team} \in [11, 13, 15, 17, 19, 21, 23, 25, 27],
\end{equation}
to quantify the strength of each team.
We randomly select a pair of teams, sample the points scored by each team's Poisson distribution, and update the Elo ratings of both teams using a small update factor $\kappa = 0.005$.
The process is then repeated a large number of times to relax each team's Elo rating to its equilibrium value.

\begin{figure}
  \includegraphics{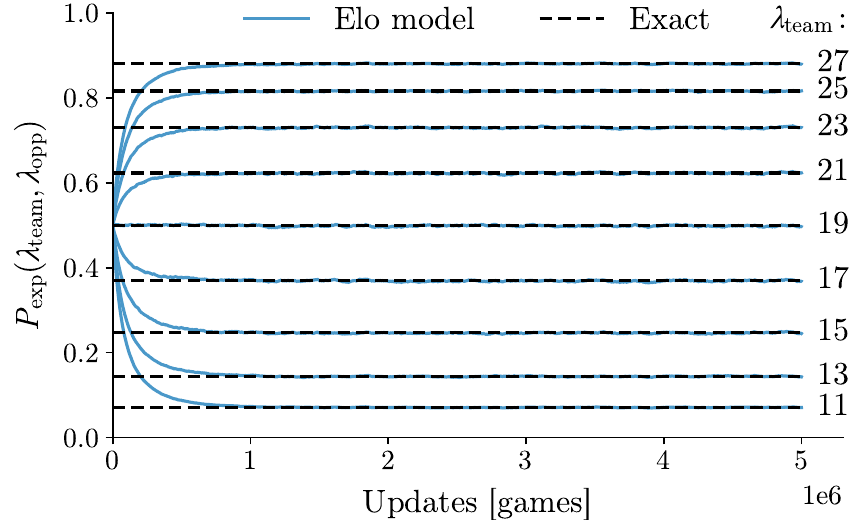}
  \caption{\label{fig:toy_model} Time series of Elo win rate predictions for nine teams \eqref{teams} in a toy-model league where each team samples $k$ points from a Poisson distribution $P(k; \lambda)$ with constant mean $\lambda_\text{team}$, annotated on the figure. The team Elo ratings are updated iteratively by selecting random matchups, sampling points from each team's Poisson distribution, and updating the Elo ratings of both teams based on the outcome of the match. The figure shows each team's predicted win rate against the league average opponent with $\lambda_\text{opp} = 19$ calculated from Eqn.~\eqref{win_prob} (solid lines). The model predictions are compared to exact win rates estimated from direct sample evaluation (dashed lines). All teams start with equal Elo rating ${\R_\text{init} = 1500}$.
  }
\end{figure}

The Elo histories are then used to calculate the expected probability $P_\text{exp}$ in Eqn.~\eqref{win_prob} that each team beats the league average opponent, described by a Poisson distribution with mean $\lambda_\text{opp}=19$, according to both teams' instantaneous Elo ratings.
Figure~\ref{fig:toy_model} plots the resulting Elo predicted win probabilities (solid lines) for each matchup and compares them to their corresponding exact win rates (dashed lines) determined by direct sample evaluation.
We find, for all practical purposes, perfect agreement between the Elo model and exact result.

This procedure is naturally overkill for an idealized model with nine teams and five million matches played. It would be far easier to simply estimate the win probability for each pair of teams directly based on the outcome of their previous matchups.
The advantage of Elo ratings is that they do not require an extensive record of games played directly between two teams to predict the probability that either team wins a future matchup. This is because Elo ratings leverage posterior information from \emph{every} match played and not just matches against the opponent of interest.
This makes Elo rating predictions ideally suited for player-based competitive games where match statistics are limited.

\section{Margin-of-victory Elo}

The original Elo model was first applied to chess rankings where there is a winner and a loser but no score.
It can be used to predict the odds with which a player beats his opponent \eqref{win_prob}, and correspondingly---by calculating various outcomes---the odds with which a player advances through a tournament. 

The model is easily extended to point based games, as demonstrated in the previous example, by ignoring the margin-of-victory and only considering wins and losses.
One shortcoming of this approach is that it cannot predict the mean or median point spreads which are often used to set betting lines and which are interesting in their own right.
Methods have been proposed to retain margin-of-victory information by allowing the parameter $\kappa$ in Eqn.~\eqref{elo} to vary with the point spread.
The idea is that if a team finds itself in a rating slump, subsequent dominant victories should raise the team's rating faster than successive close wins.
The method appeared in world football Elo ratings \cite{footballratings, elo_blog} and was adapted to forecast NFL and NBA team performance on the analytics and news platform \href{www.fivethirtyeight.com}{FiveThirtyEight} \cite{538NFL, 538NBA}.

Spread-dependent $\kappa$-factors possess some desirable features, but they also introduce a number of problems.
There is no obvious form for a margin-of-victory multiplier on $\kappa$, and introducing one still does not get one any closer to predicting the mean or median score of a game.
There are also issues of autocorrelation, i.e.\ great teams should not be over rewarded for winning by large margins because that's what great teams do \cite{538NFL}.

There is, alternatively, a more natural extension of Elo ratings which allows one to incorporate margin-of-victory information and predict the mean and median score of future games.
We start by emphasizing that Elo ratings only require agents which compete to win games and that ``winning'' is a subjective definition itself.
Let $p_s$ and $p_a$ denote the points scored and points allowed by each team respectively.
The usual win criteria is defined as
\begin{equation}
  \text{win}: p_s - p_a > 0.
\end{equation}
This is the fair or democratic criteria as it gives each competitor a level playing field.
Just as easily, one could define winning as
\begin{equation}
  \text{win}: p_s - p_a > n,
\end{equation}
where $n$ is a finite margin-of-victory line.
This is equivalent to a normal Elo rating system where a point handicap is applied against one team.

To calculate the probability that a team beats an opponent by more than $n$ points, one need only calculate the Elo ratings of the team and their opponent under an $n$-point handicap. i.e.\ according to the win criteria
\begin{equation}
  \text{win}: p_s - p_a - n > 0.
\end{equation}
The handicapping procedure splits team ratings into two different groups: a rating handicapped by $n$ points, and a rating advantaged by $n$ points.

To get a better feel for what this does, imagine an average team under the classical Elo system which corresponds to a handicap $n=0$. 
As we apply a nonzero handicap $n$ to the home team, it becomes increasingly difficult for the team to overcome their handicap and win games.
The team's handicapped Elo rating will drop, and eventually will fall below even the worst non-handicapped team.
Simultaneously, we must also keep track of each team's performance \emph{against} handicapped teams, i.e.\ for $n \rightarrow -n$.
Naturally, this will have an opposite effect, and each team's advantaged rating will rise as their advantage increases.

The procedure thus creates mirror copies of each team's rating at each value of the spread: the handicapped rating $\R_\text{team}(n)$, and advantaged rating $\R_\text{team}(-n)$.
To predict the probability that a team will beat their opponent by more than $n$ points, we calculate the difference between the handicapped and advantaged ratings
\begin{equation}
  \label{rtg_diff}
  \Delta\R = \R_\text{team}(n) - \R_\text{opp}(-n)
\end{equation}
and use Eqn.~\eqref{win_prob} to calculate $P_\text{exp}(\Delta \R)$ as before.

The procedure can easily be repeated for each value of $n$ spanning a large range of spreads to calculate the margin-dependent Elo rating for every reasonable margin-of-victory.
The ratings can then be used to predict the probability that a team beats its opponent by \emph{any} point number.
The procedure simply constructs the point spread's cumulative distribution function (CDF)
\begin{equation}
  \label{cdf}
  F(n) = \sum_{s=n}^\infty P(s),
\end{equation}
where $s = p_s - p_a$.
Equation~\eqref{cdf} encodes the full probability distribution of allowable point spreads and can be used to calculate both mean and median point values.

\begin{figure}
  \includegraphics{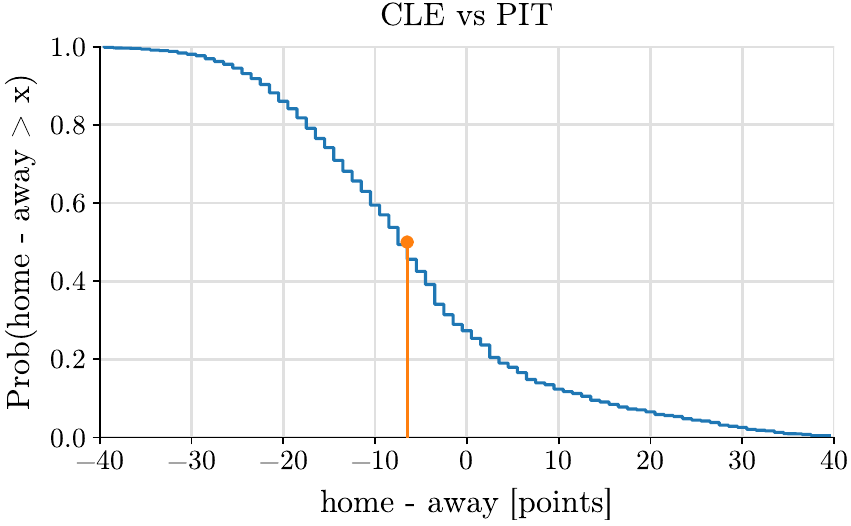}
  \caption{\label{fig:example} Example cumulative probability distribution function (CDF) of point spreads calculated for a specific matchup from the margin-dependent Elo model (blue line). The distribution's median value is marked by a vertical orange line.}
\end{figure}

Figure~\ref{fig:example} shows, for illustrative purposes, the resulting cumulative probability distribution calculated from Eqn.~\eqref{cdf} for an NFL game in which the Cleveland Browns host the Pittsburgh Steelers. 
Note that the sign convention used here is opposite what is typically used in Vegas; negative spreads in Fig.~\ref{fig:example} mean Cleveland is an underdog while positive spreads mean they are favored.
The distribution median, indicated by a vertical orange line, shows that the Browns are significant underdogs at home.

\section{Descriptive Statistics}

A primary advantage of margin-dependent Elo ratings is that they predict the full probability distribution of expected outcomes.
This allows one to calculate various statistical properties of the distribution such as its mean, median and standard deviation.
It also allows one to calculate the probability that the observed point spread falls above or below arbitrary values as well as within specific intervals---probabilistic outcomes which are commonly used in betting circles.

\subsection{Median point spread}

The median point spread is the value which gives both teams equal odds to cover.
It's a popular betting statistic as it generally promotes equal numbers of bets on both sides of the line which increases betting volume and minimizes risk for the house.
In the margin-dependent Elo model, we can easily estimate the median point spread by calculating the handicapped and advantaged ratings at each value of the spread, and by finding the spread where the win probability between the handicapped and advantaged teams is closest to 50\%.
This is equivalent to the point spread $\tilde{s}$ such that
\begin{equation}
  \R_\text{team}(n) \approx \R_\text{opp}(-n).
\end{equation}

One should be reasonably concerned that the handicapped Elo ratings will be unreliable for large values of the spread as the handicapped team will seldom win.
Indeed, there are large uncertainties in predicting a blowout as the statistics are simply insufficient to extrapolate from prior to future events.
In contrast to these extreme spread values, the spread median is---by definition---the value where wins and losses occur with equal probability and hence events are maximally frequent to maintain rating equilibrium.
One should expect the median values to be relatively accurate within the limitations of the model.

\subsection{Mean point spread}

Although the mean point spread is less commonly used in betting circles, it's nevertheless interesting to compute.
The mean point spread can be obtained from the cumulative point spread distribution $F$ via integration by parts
\begin{equation}
  \label{parts}
  \bar{s} = \sum\limits_{s_\text{min}}^{s_\text{max}} s P(s) = s_\text{max} - \sum\limits_{s_\text{min}}^{s_\text{max}} F(s),
\end{equation}
where the boundary term in Eqn.~\eqref{parts} simplifies since the CDF converges at the point spread extrema,
\begin{equation}
  \lim F(s)=
  \begin{cases}
    0, & s \to +\infty,\\
    1, & s \to -\infty.
  \end{cases}
\end{equation}

The mean and median are, in general, equal only for symmetric distributions and may differ significantly in lopsided matchups where Elo differences are large.
Such differences are particularly informative when predicting spread outcomes away from the median expectation value, e.g.\ when predicting a blowout. 

\section{Orthogonal point combinations}

\begin{figure}
  \includegraphics{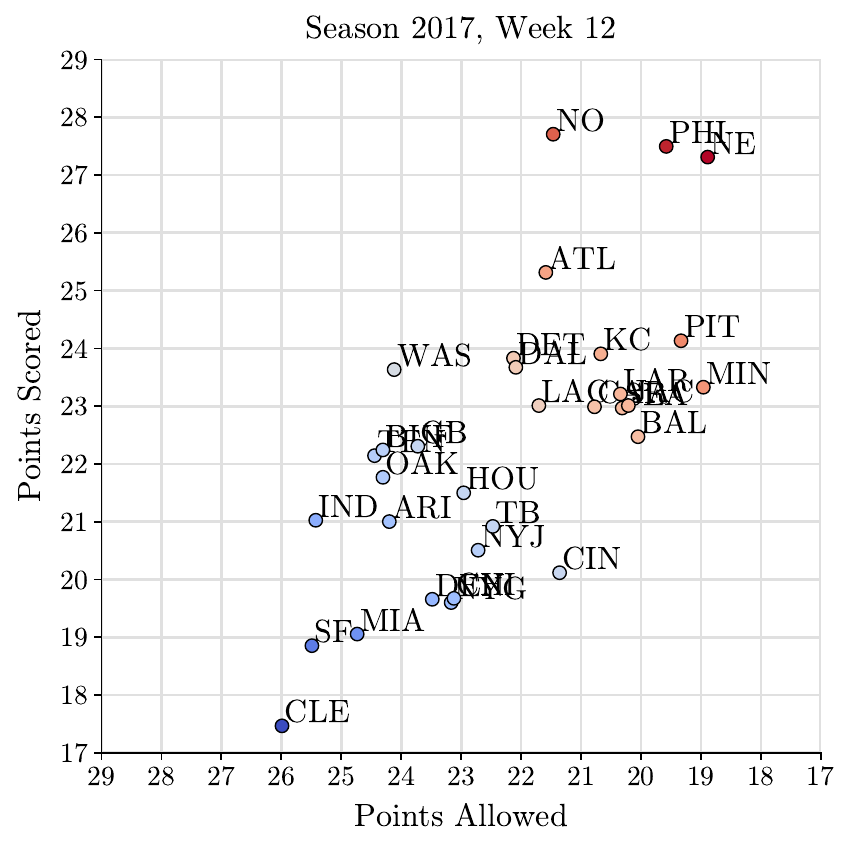}
  \caption{\label{fig:ratings} Predicted mean points scored and mean points allowed against a league average opponent calculated for each team in week 12 of the 2017 NFL season using the margin-dependent Elo model.}
\end{figure}

In addition to the point spread $s = p_s - p_a$, it's also possible to predict the distribution of \emph{total} points scored $t = p_s + p_a$.
The point spread and point total values represent symmetric and asymmetric components of the two teams' point distributions and form an orthogonal basis for the game score. 

Point total ratings can be constructed in an analogous fashion to the aforementioned point spread ratings by using an Elo win criteria with the modified form
\begin{equation}
  \label{win_total}
  \text{win}: p_s + p_a - m > 0,
\end{equation}
where $m$ applies a handicap to the point total $t$.
This is equivalent to the margin-dependent spread condition, except that we've flipped the sign of the opponent's points.
In doing so, the spread difference turns into a sum, and a team ``wins'' their game if its score exceeds its opponent's \emph{negated} score by a finite margin of victory, i.e.\ if the point total of both teams exceeds a given threshold.

For example, we might choose to flip the sign of all away teams.
We then take the difference between every home team score and every (negative) away team score, equal to both teams' combined point total.
If this difference exceeds a pre-specified point total handicap, we say the home team wins and away team loses or \emph{vice versa}.
Naturally, the procedure is then repeated but with the sign flip applied to all home teams such that the arbitrary win criteria is reflected, i.e.\ with the logic negated.

The procedure consequently splits each team's rating into two mirror copies as before: one for both positive and negative values of the handicap $m$.
Games which rack up a large number of points increase ratings with positive handicap values $m > 0$ and decrease ratings with negative handicap values $m < 0$.
Since model predictions always pair a positive handicap value with a negative handicap value, e.g.\ $\mathcal{R}_\text{team}(m)$ with $\mathcal{R}_\text{opp}(-m)$, such high scoring games \emph{increase} the gap between ratings with positive and negative handicaps, and subsequently increase the probability $P_\text{exp}(t > m)$ that the two teams would exceed the specified point total in subsequent matchups.

Once the point spread and point total Elo ratings are determined for two teams, it is possible to predict not just the game's point spread and point total, but also each team's points scored.
We observe that the points scored and points allowed by each team are just orthogonal combinations of the point spread and point total,
\begin{align}
  \bar{p}_s &= \tfrac{1}{2}(\bar{s} + \bar{t}),\nonumber \\
  \bar{p}_a &= \tfrac{1}{2}(\bar{s} - \bar{t}),
  \label{points}
\end{align}
where $\bar{s}$ is the average point spread and $\bar{t}$ the average point total.

For example, one interesting application is to use Eqn.~\eqref{points} to predict the points scored and points allowed by every team against a league average opponent.
Scatter plotting each team's points scored and points allowed will then give us an indication of which teams are strong offensively (score many points) and which teams are strong defensively (allow few points).

We demonstrate such a calculation in Fig.~\ref{fig:ratings} where we show the projected points scored and points allowed for each team against a league average opponent in week 12 of the 2017 NFL season.
Note that higher points scored and lower points allowed tend to correlate with stronger offenses and defenses respectively, but the correspondence is not one-to-one as a defense may score points and an offense may give up points.

\section{Applying the model to the NFL}

In this section we explain in detail our application of the aforementioned margin-dependent Elo rating system to the National Football League (NFL) using game data from 2009--2017.
The purpose of this section is to demonstrate the efficacy of the proposed model in a real world scenario and to give examples of specific model calculations which are not possible using existing Elo methods.
We also explain several modifications which were necessary to tailor the margin dependent Elo model to the problem at hand.

We emphasize that the present model cannot be used to ``beat Vegas'', at least in the usual sense, as we do not account for weather, injuries or personnel changes which contribute significantly to NFL betting lines.
Nevertheless, the point spread and point total distributions calculated from the margin-dependent Elo model are surprisingly consistent with opening betting lines and contain a wealth of novel information.

\subsection{Initializing the Elo ratings}

\begin{figure}
  \includegraphics{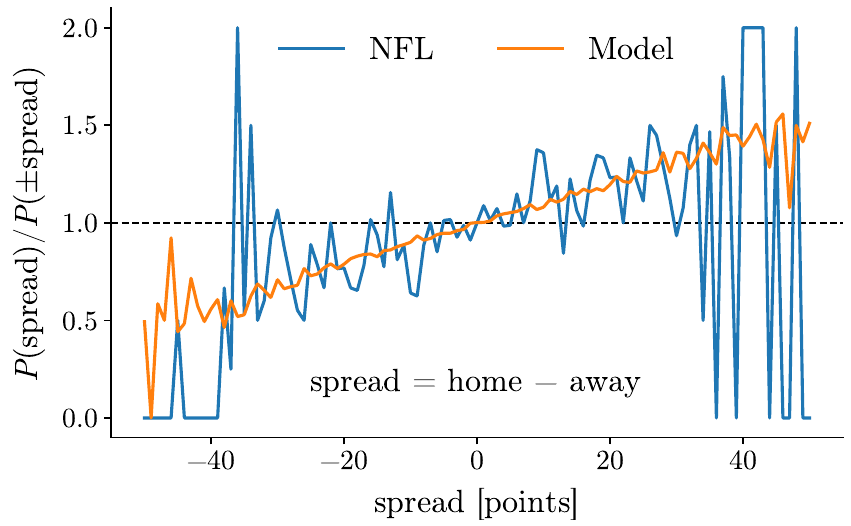}
  \caption{\label{fig:home_field} Ratio of probabilities of observing a home field biased point spread over an unbiased point spread as a function of point spread. The blue line denotes calculations derived from NFL game scores and the orange line model predictions from margin-dependent Elo ratings. Model predictions are drawn from samples of the model's predicted point spread distribution for each game. The effect of no home field advantage is indicated by a black dashed line for reference.}
\end{figure}

\begin{figure*}
  \includegraphics{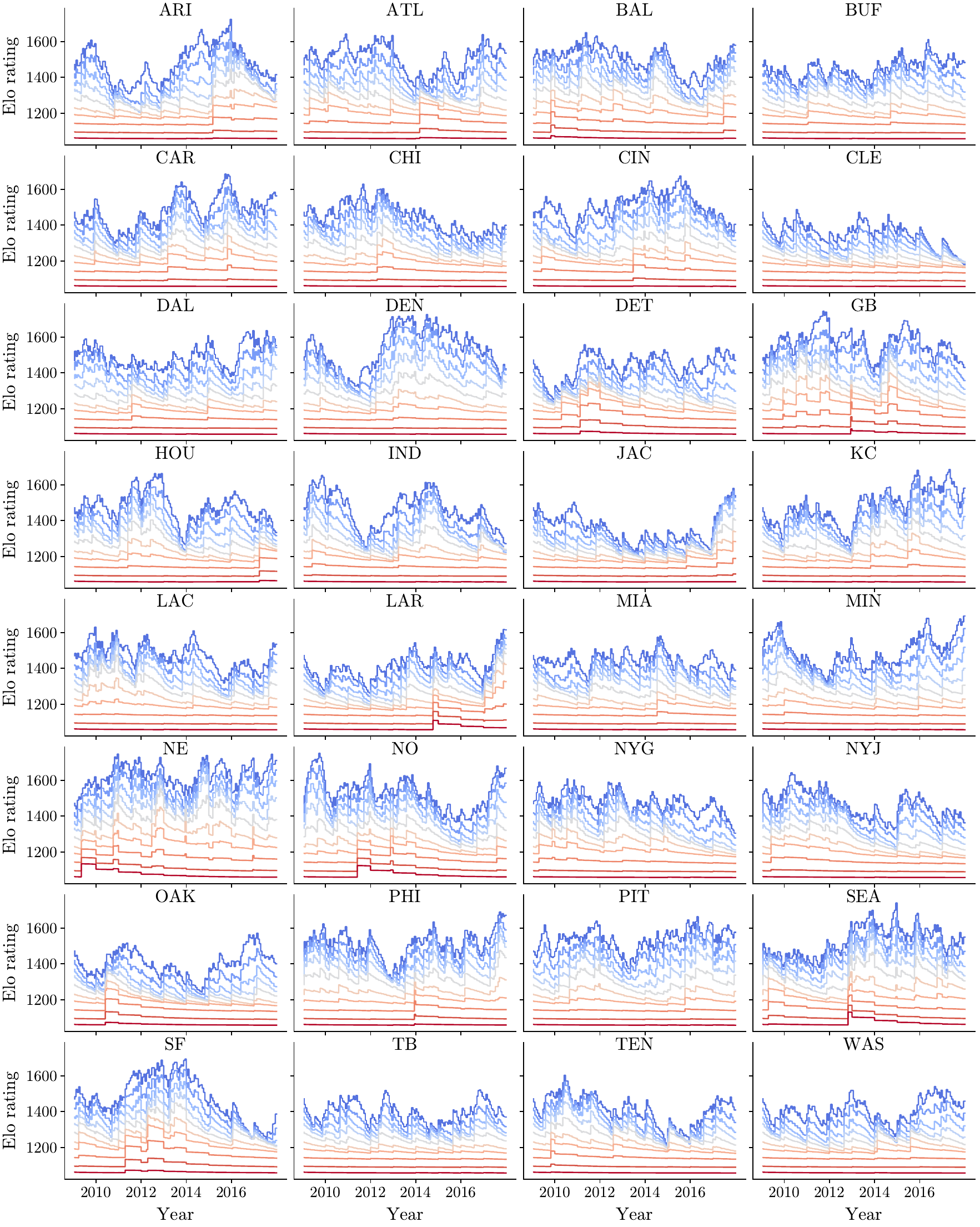}
  \caption{\label{fig:history} Handicapped NFL team Elo ratings 2009--2017. Colored lines denote different handicaps. Dark blue denotes a handicap of 0 points and dark red a handicap of 50 points. Intermediate colors are constant handicap increments of 5 points. Ratings for Rams and Chargers persist through relocation. Not shown: advantaged ratings which mirror handicapped ratings.}
\end{figure*}

In the original Elo model, it is customary to initialize teams to the same rating such that $P_\text{exp} = 0.5$ for every team's first match.
Naturally, the existence of handicaps in the margin-dependent Elo model complicates matters since handicapped matches are no longer fair, i.e.\ we shouldn't expect a generic handicapped team to win with 50--50 odds.
The correct, more general, starting assumption is to assume that teams sample points from minimum bias distributions, or equivalently, that teams are interchangeable and sample scores generically.

We enforce this starting criteria by calculating the minimum bias distribution of point spreads and point totals from historical NFL games and using them to derive starting Elo rating values which reproduce the observed distributions.
For example, the formula for initializing the point spread Elo ratings can be determined by plugging Eqn.~\eqref{rtg_diff} into Eqn.~\eqref{win_prob} and inverting to find $\Delta \R(n)$ as a function of $P_\text{exp}$,
\begin{equation}
  \Delta \R(n) = \sqrt{2} \sigma\,  \erf^{-1}(2 P_\text{exp} - 1).
\end{equation}
The initial spread-dependent Elo rating $\R_\text{init}(n)$ is then given by
\begin{equation}
  \R_\text{init}(\pm n) = \R_0 \mp \frac{\sqrt{2}\sigma}{2} \erf^{-1}(2 P_\text{exp} - 1),
\end{equation}
where $\mathcal{R}_0 = 1500$ is the customary starting value in the original Elo model (handicap of zero) and ${P=P_\text{exp}(s > n)}$ is the probability that the minimum bias point spread $s$ is greater than $n$ points.
The proper starting Elo ratings for the point total can be furnished in an analogous fashion.

\subsection{Regressing towards the mean}

Elo ratings are updated using posterior information which registers the moment a game is completed.
Clearly, current Elo ratings cannot reliably predict team performance far into the future, say 10 years from now.
In fact, such predictions would likely yield little or no information at all.
The information in the Elo ratings begins to decay and regress towards the mean the moment they are updated after a game.
This effect exists between subsequent games in the regular season, but it is largely negligible compared the much larger lapse in time between individual seasons.
To better account for hard to quantify changes that occur during the offseason, we follow the method used by FiveThirtyEight and regress each team rating toward its mean handicap value by a fixed percentage each offseason \cite{538NFL, 538NBA}.
We find that regressing each rating to 60\% of its starting value each season for point spreads and 70\% for point totals leads to the best description of game outcomes.

\subsection{Home field advantage}

\begin{figure}
    \includegraphics{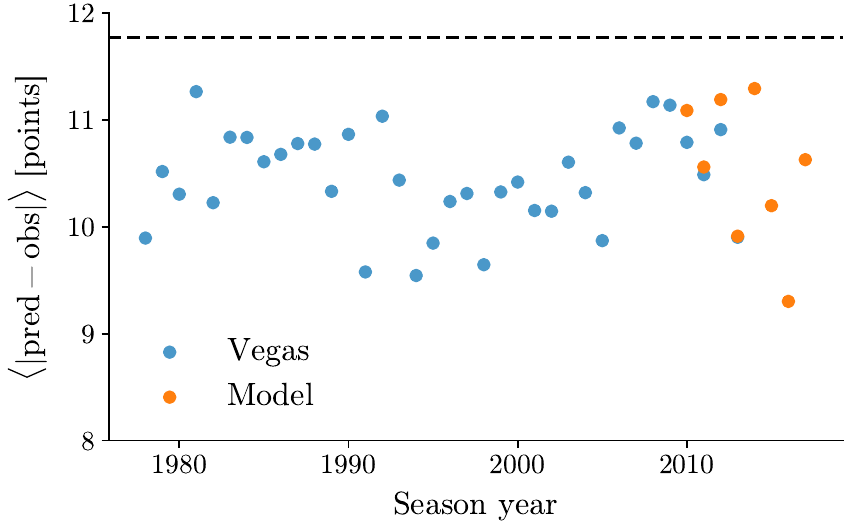}
    \caption{\label{fig:mean_abs_error} Mean absolute error of NFL point spread residuals calculated for years 1979--2017. Blue points are Vegas betting lines from Ref.~\cite{vegaslines}, and orange points are calculations from the margin-dependent Elo model. The horizontal dashed line is the mean absolute error of a ``dumb'' model which predicts fixed spread of zero for each game.}
\end{figure}

Team's generally perform better at home and worse on the road.
Football analysts often quote a ${\sim}\,3$ point advantage for the home team relative to what the spread would otherwise be on a neutral field. 
This tacitly assumes that home field advantage applies a translational shift to the game's spread distribution, when in reality, a shift is just one of a number of reasonable transformations.
For instance, a skewness transformation seems equally plausible at face value.

The margin-dependent Elo model admits an exceptionally simple and intuitive method to incorporate home field advantage, previously implemented in paired comparison models \cite{glickman_thesis}, and adapted for this work.
One can think of home field advantage as a temporary boost to a team's Elo rating, i.e.\ teams with home field advantage play like slightly better versions of themselves.  
Home field advantage can thus be incorporated by increasing the Elo gap between home teams and away teams by a constant offset $\mathcal{R}_\text{hfa}$ which we assume to be equal at every value of the handicap, i.e.\
\begin{equation}
  \label{hfa}
  \Delta \mathcal{R} = \mathcal{R}_\text{home} - \mathcal{R}_\text{away} + \mathcal{R}_\text{hfa}.
\end{equation}
This means, for instance, that losses which occur on the road are penalized less heavily than losses which occur at home since they are modeled as if the opponent had a slightly higher Elo rating.
Additionally, when predicting the outcome of future matches, one must also add $\mathcal{R}_\text{hfa}$ to the matchup's Elo difference to account for the home team's advantage.

We can quantify the effect of home field advantage by histogramming the home-field biased and unbiased point spreads,
\begin{align}
  s_\text{biased} &= \{p_\text{home} - p_\text{away}\},\nonumber \\
  s_\text{unbiased} &= \{p_\text{home} - p_\text{away}\} \cup \{p_\text{away} - p_\text{home}\},
\end{align}
and normalizing each set of binned counts before dividing both histograms.
We plot this ratio (blue line) in Fig.~\ref{fig:home_field}, and compare its value to the equivalent quantity derived from model predictions (orange line) with ${\mathcal{R}_\text{hfa} = 54}$~points.
The model predictions with a home field advantage correction implemented according to Eqn.~\eqref{hfa} are in good agreement with the data and illustrate the robustness of the present implementation. 
Moreover, we demonstrate that the home field advantage implemented in this work has the desired effect on the entire point spread distribution, not just its average value. 

\subsection{Validation}

\begin{figure}
  \includegraphics{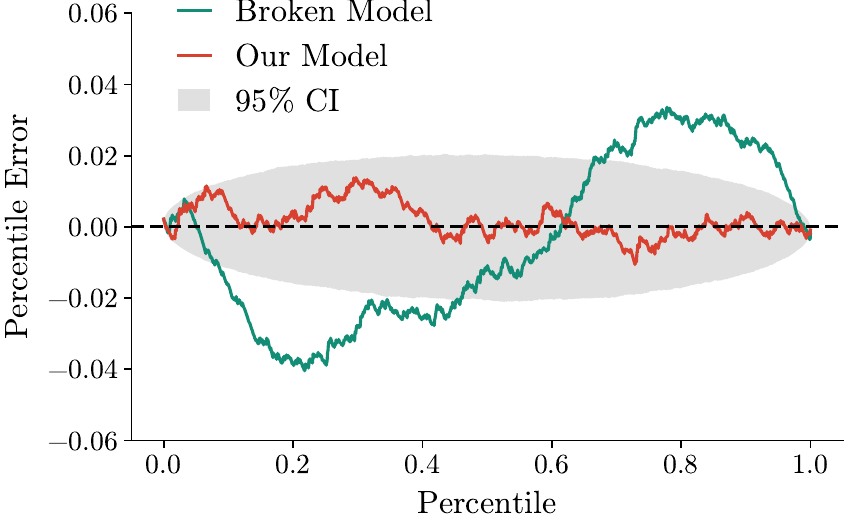}
  \caption{\label{fig:percentiles} Difference between the list of ordered percentiles obtained from back-testing model predictions, and the list of ordered percentiles obtained from sampling a uniform distribution (red line).
For accurate models which correctly estimate their own uncertainties, 95\% of the percentile errors should fall within the gray band, calculated by comparing different samples from a uniform distribution.
  We also show, for reference purposes, the same percentile errors calculated from a broken model where the Elo model predictions are matched with random games (green line).
  }
\end{figure}

\begin{figure*}
  \includegraphics{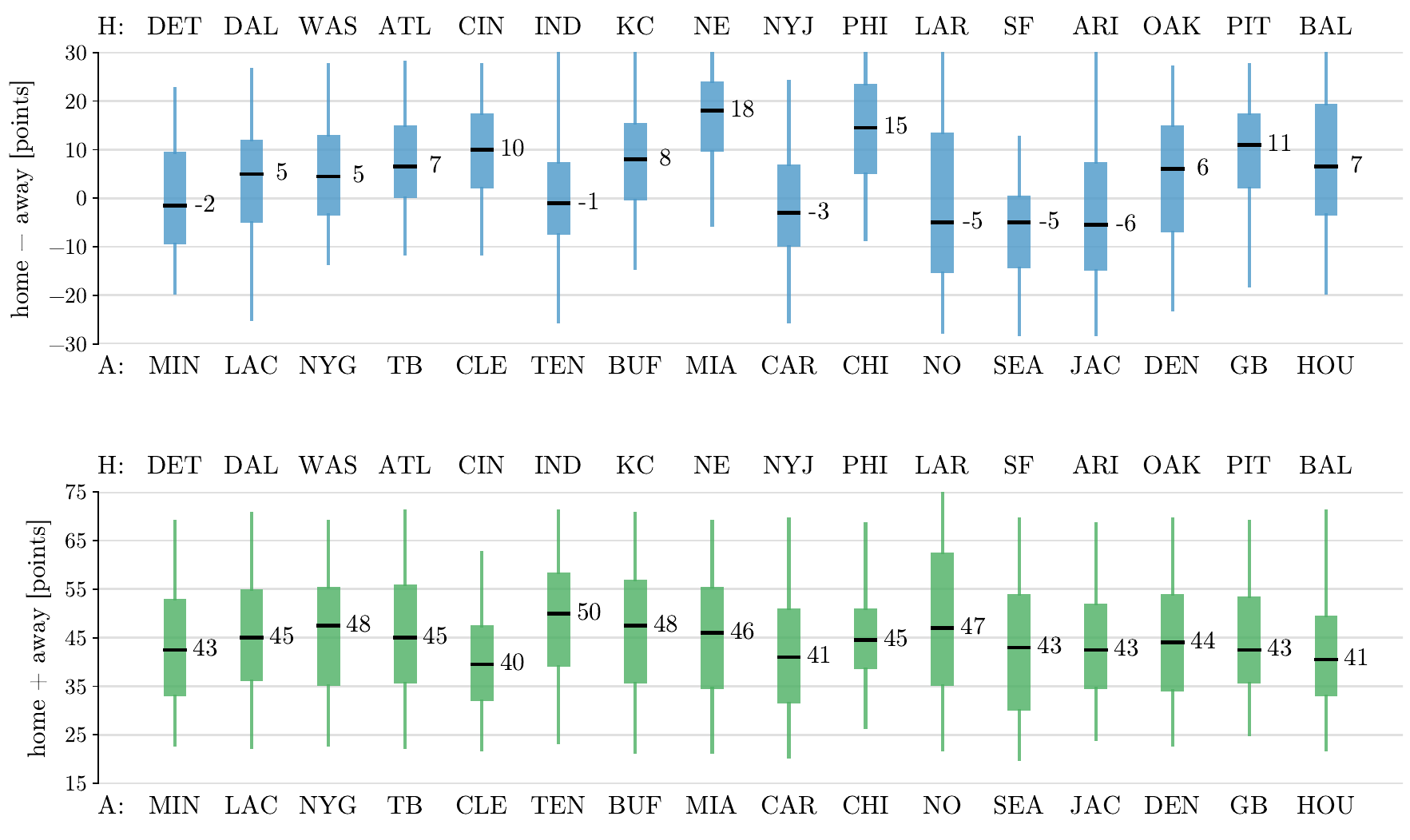}
  \caption{\label{fig:predict} NFL score predictions determined from the margin-dependent Elo model for the twelvth week of the 2017 season. Black dashes are the prediction medians, boxes are the 25--75\% inner-quartile range, and whiskers cover 90\% of the distribution tails. Home field advantage was accounted for in the spreads by handicapping the away team $\Delta\R_\text{hfa}=54$ Elo rating points.}
\end{figure*}

The margin-dependent Elo model constructs a time series of Elo ratings for each team and for each value of the point spread or point total.
Figure~\ref{fig:history} shows, for illustration purposes, the handicapped Elo point spread ratings obtained from the model once it has been applied to the NFL.
These ratings, once furnished, can then be used to predict point spreads and point totals for any game, using the information that was available at the time.

The model predicts a \emph{distribution} of point spreads and point totals for each game so it is important to distinguish between the model's accuracy and its precision. 
For example, a model which predicts every spread to be less than a million is accurate but imprecise.
Conversely, a model which predicts every spread to be one point higher than its observed value is precise but inaccurate.
In general, probabilistic models should self report errors that are sufficiently large to maintain accuracy but no larger.

One commonly used measure to assess non-probabilistic betting models which predict only a single ``best guess'' value for each game is the mean absolute error, defined as the arithmetic mean of absolute model residuals.
Figure~\ref{fig:mean_abs_error} compares our model's mean absolute error against available Vegas betting lines for years where either data is available\footnote{The \texttt{nfldb} database includes years 2009--present. The Vegas lines are for years 1978--2013.}.
Although our source of historical NFL betting lines is limited, we clearly see that the model residuals are of similar magnitude to Vegas lines, albeit slightly worse for years which overlap.

There is, of course, additional information in the margin-dependent Elo model that is not evaluated by the mean absolute error.
For example, it's important to test that the model correctly predicts the frequency of upsets and blowouts, rare events which fall in the tails of the predicted spread distribution.
Simply put, if we predict an event to occur frequently, then we should observe it to occur frequently, and if we predict an event to occur rarely, then we should observe it to occur rarely. 
More rigorously speaking, if nature samples a point spread $s$, we can determine the percentile of the sampled spread \emph{ex post facto} given our's model's prediction for the cumulative distribution $F(s)$.
We can then check that the percentiles of the postdicted events sample a unit random variable.

Typically, this is accomplished using what's known as a quantile-quantile plot, where the model percentiles are scatter plotted against a uniform partition from 0 to 1.
In this work, we choose a slightly different visualization method and plot in Fig.~\ref{fig:percentiles} the \emph{difference} between the ordered model percentiles and the percentiles drawn from a uniform partition of the CDF.
If the present model were perfectly accurate, its percentile samples would be that of a uniform random variable which fall within the figure's gray band 95\% of the time.

We see that the actual model calculations (red line) generally fall within this confidence interval and hence are quite good.
This suggests that the margin-dependent Elo model correctly estimates it's own uncertainty when predicting the distribution of future point spreads.
Moreover, we can see that the agreement is largely nontrivial by breaking the model so that the model predictions are matched with randomly selected point spreads (green line).
In this case, the ``broken'' model predictions largely fall outside the 95\% confidence interval which indicates poor accuracy.
This indicates that the present model predictions are not just accurate---the obtained accuracy is also nontrivial.

\subsection{Application}

The margin-dependent Elo model predicts probability distributions for the outcomes of games, either point spreads or point totals, which can easily be visualized as box plots, violin plots or through similar visualization methods.
The advantage of presenting the information as a probability distribution, is that it emphasizes the noisy and highly uncertain nature of NFL games, information which is not conveyed by a single Vegas point spread or point total.

We apply the present model to week 12 of the 2017 NFL season in Fig.~\ref{fig:predict} to illustrate the probabilistic nature of the model.
Interestingly, both the mean and width of the distributions vary from game to game which indicates that some games outcomes are less certain than others.

It is of course possible to use these distributions to predict the probability that a game's point spread (or total) falls above or below a given line, and hence estimate a bet's expected return on investment assuming perfect model accuracy.
We find however, that in practice such a betting scheme will not make one rich as it neglects significant sources of external information which affect the game's expected outcome.
For example, blithely applying the present model to predict the point total of a snowy game in Detroit would likely over estimate the points scored.

\section{Concluding remarks}

In this work, we formulated a simple extension to the original Elo model which generalizes the algorithm from binary win-loss predictions to full probability distributions of point spread and point total outcomes.
The generalized model allows us to predict the median and mean point scores, as well as any other statistical quantity which can be derived from the underlying point distributions, for instance the interquartile ranges. 

We have applied the model to the NFL for the years 2009--2017 and shown how the algorithm can be used to generate predictions for each game.
Moreover, we explain how the point spread and point total predictions can be combined to predict the points each team would score and allow against a league average opponent. 

The model is back tested to assess the distribution of model residuals.
Our findings indicate the probabilistic predictions are highly accurate with a mean absolute error that is comparable to, albeit slightly worse than, Vegas betting lines.
Naturally, the model's accuracy would benefit from additional factors such as personnel changes, weather and extraneous circumstances which are already accounted for in betting lines.

We emphasize that the present algorithm should readily generalize to any point based competitive game.
For example, the model would be ideally suited for the National Basketball Association (NBA) due to its high scoring statistics and frequent game play. 
We leave these applications to future studies.

\medskip

All software used in this work is open source.
The \texttt{melo} Python package is publicly available at \mbox{\url{github.com/morelandjs/melo}}.
Several third-party software packages were also invaluable: the \texttt{nfldb} Python package written by Andrew Gallant \cite{nfldb}, the \texttt{scipy} library for scientific computing \cite{scipy}, and the \texttt{scikit-optimize} package for optimizing noisy and expensive functions \cite{skopt}. 

\bibliography{marginelo}

\end{document}